\documentclass[a4paper,11pt]{article}
\pdfoutput=1 

\usepackage{jheppub} 

\usepackage[T1]{fontenc} 

\title{\boldmath On searches for anomalous WWW interaction at the LHC}

 \author{B.A. Arbuzov}
 \author{I.V. Zaitsev}
\affiliation{M.V. Lomonosov Moscow State University,\\119991 Moscow, Russia}

\emailAdd{arbuzov@theory.sinp.msu.ru}
\emailAdd{zaitsev@theory.sinp.msu.ru}

\abstract{The process $p+p\to W^\pm W^+ W^-$ with lepton decays of $W$-bosons is shown to be promising
for improving conditions for studies of parameter $\lambda$, which is the  anomalous
triple $W$-bosons interaction coupling constant.}

\begin{document}
\maketitle
\flushbottom

\section{Introduction}
\label{sec:intro}

The totality of data acquired by LHC confirms predictions of the Standard Model.
Meanwhile the numerous attempts for obtaining even indications for deviations from the SM
were not successful. Mostly these efforts were aimed at a quest for effects of the so-called New Physics, that is of wouldbe effects of the Supersymmetry, Extra Dimensions, Dark Matter and some other
radical hypotheses. However, there may be also quite interesting wouldbe effects, which are connected just with more conventional physics of the Standard Model, but with those, which are of a   non-perturbative origin.

Indeed, the Standard Model has two main constituents: QCD and the electroweak  theory (EWT in what follows). Both theories are the renormalizable ones. Hence the perturbation theory is duly defined
and give quite definite predictions.

Nevertheless, in QCD the existence of the non-perturbative contributions is inevitable.
They define {\it e.g.} non-zero vacuum averages: gluon condensate and quark condensate
\begin{equation}
<G^{\mu \nu}\,G_{\mu \nu}>,\quad  <\bar q\,q>;  \label{cond}
\end{equation}
the behaviour of running coupling $\alpha_s(Q^2)$ and some other actual effects.
For the moment the existence of non-perturbative effects in EWT is not evident
to the same extent as in QCD. However, the arguments were expressed for such effects being present
in the electroweak interactions as well.

We would discuss in the present note an example of the would be non-perturbative efffect,
connected with anomalous triple electro-weak bosons
interaction, first proposed in works~\cite{Hag1,Hag2}.
\begin{eqnarray}
& &-\,\frac{G}{3!}\,F\,\epsilon_{abc}\,W_{\mu\nu}^a\,W_{\nu\rho}^b\,
W_{\rho\mu}^c\,;
\label{FFF}\\
& &W_{\mu\nu}^a\,=\,
\partial_\mu W_\nu^a - \partial_\nu W_\mu^a\,+g\,\epsilon_{abc}
W_\mu^b W_\nu^c\,.\nonumber
\end{eqnarray}
with  form-factor $F(p_i)$, which guarantees effective interaction~(\ref{FFF}) acting in a limited region of the momentum space. A possibility of the spontaneous generation of effective interaction~(\ref{FFF}) was first considered in work~\cite{BAA92} and was studied in more details
in work~\cite{AZPR}.
Interaction constant $G$ is connected with
conventional definitions in the following way
\begin{equation}
G\,=\,-\,\frac{g\,\lambda}{M_W^2}\,;\label{Glam}
\end{equation}
where $g(M_Z) \simeq 0.65$ is the electroweak coupling.
Interaction~(\ref{FFF}) contains both anomalous three-boson vertex and four-boson vertex, which are proportional to constant G~(\ref{Glam}) (five-boson and six-boson terms will not enter to the present calculations).
The best experimental limitations obtained in experiments at LHC for parameter $\lambda$ read
\begin{eqnarray}
& &-0.011<\lambda<  0.011\,; \label{lambda17}\\
& &-0.0088<\lambda<  0.0095\,; \label{lambda19}\\
& &-0.0099<\lambda<0.0104 \,. \label{lambdaWZ}
\end{eqnarray}
Result~(\ref{lambda17}) corresponds to LHC energy $\sqrt{s} = 8\,TeV$ and integral luminosity
$L = 18\,fb^{-1}$\cite{CMS17}, while recent result~(\ref{lambda19}) corresponds to LHC energy $\sqrt{s} = 13\,TeV$ and integral luminosity
$L = 36.4\,fb^{-1}$\cite{CMS19}. We see, that the progress in an accuracy from (\ref{lambda17}) up to  (\ref{lambda19}) is not very substantial. The process, which was studied in~\cite{CMS17,CMS19}
\begin{equation}
p+p\to jet\,+\,jet\,+\,W^\pm\,.\label{jjW}
\end{equation}
Result~(\ref{lambdaWZ}) is extracted from recent work~\cite{WZCMS} in which process $p\,p\to W\,Z$
was studied at $\sqrt{s} = 13\,TeV$. The measurements of the anomalous triple gauge couplings at LHC 
using this process, as well as process $p\,p\to W\,\gamma$, were discussed in detail in recent 
work~\cite{WZTH}. 

In the present paper we would discuss another effective channel for
a study of the wouldbe interaction (\ref{FFF}).

\section{$W^+W^+W^-$ and $W^-W^-W^+$ production in $p p$ reactions at the LHC}
\label{sec:WWW}
 Let us consider no jets processes
\begin{eqnarray}
& &p+p \to W^+\,W^+\,W^-\,+\,no\,\,jet\,; \label{W+}\\
& &p+p \to W^-\,W^-\,W^+\,+\,no\,\,jet\,. \label{W-}
\end{eqnarray}
In these processes we have to study the following decays of final state $W$-bosons
\begin{eqnarray}
& &W^+\,W^+\,W^-\,\to \mu^+ \mu^+\,e^- + neutrinos;\nonumber \\
& &W^+\,W^+\,W^-\,\to e^+ e^+\,\mu^- + neutrinos;\label{W+dec}\\
& & W^-\,W^-\,W^+\,\to \mu^- \mu^-\,e^+ + neutrinos;\nonumber \\
& & W^-\,W^-\,W^+\,\to e^- e^-\,\mu^+ + neutrinos.\label{W-dec}
\end{eqnarray}
So, we have in processes~(\ref{W+},\ref{W-}) no jets, three charged leptons in the final state and a missing energy, which is carried by neutrinos.
Decays~(\ref{W+dec},\ref{W-dec}) contain leptons of the same flavour bearing same electric charges, and different
flavour leptons with opposite charges.
This signature is, practically, free from background of processes, different from basic reactions~(\ref{W+},\ref{W-}).
\begin{figure}
\includegraphics[scale=0.73]{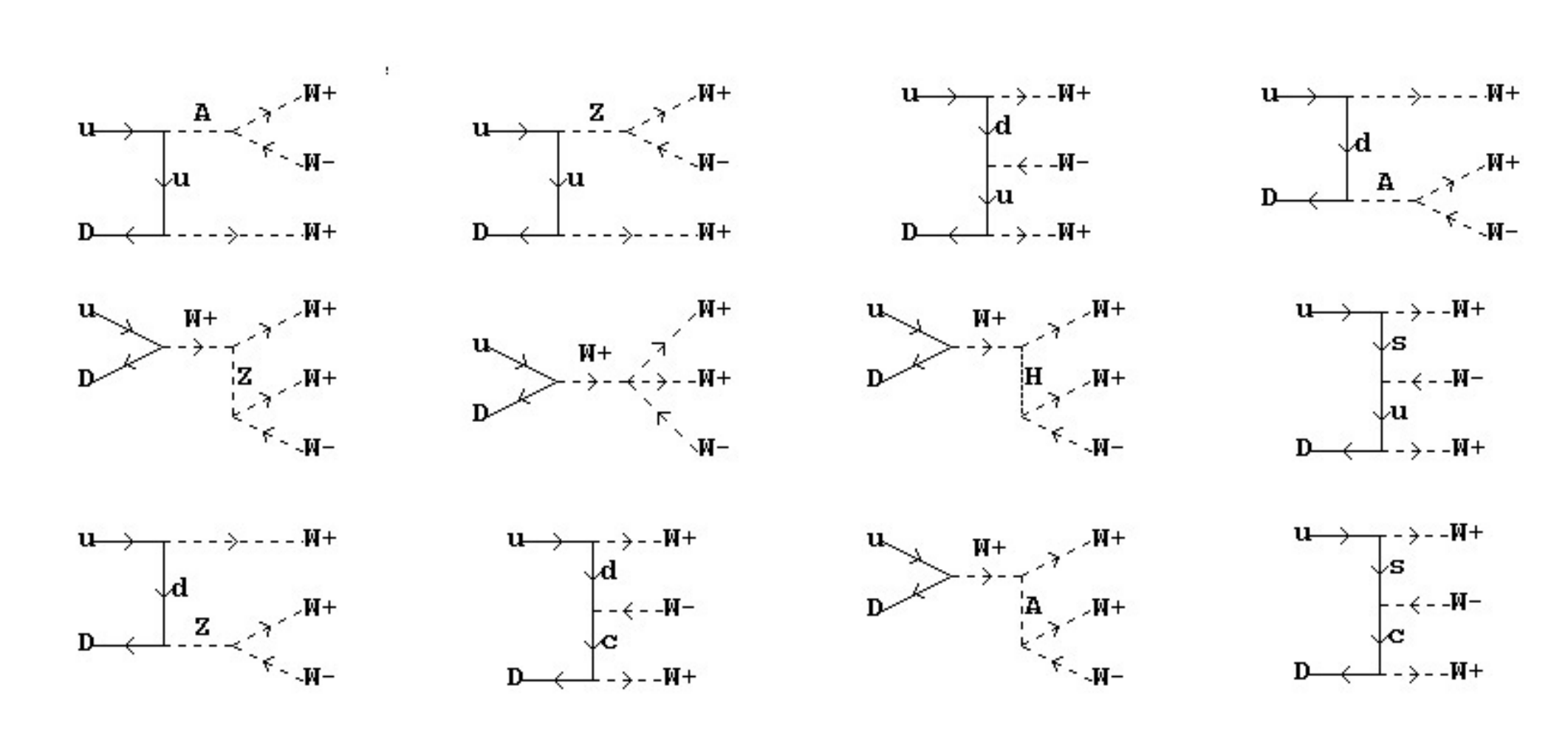}
\caption{Diagrams for subprocess $u+D \to W^+W^+W^-$.}
\label{fig:W++}
\end{figure}
On the contrary, provided we have pairs $l^+ l^-$, there are significant contributions from photons,
Z-bosons {\it etc}, and cross-sections become more than an order of magnitude larger than those for
reactions~(\ref{W+},\ref{W-}), while the effect under the study remains practically the same.

\begin{figure}
\includegraphics[scale=0.73]{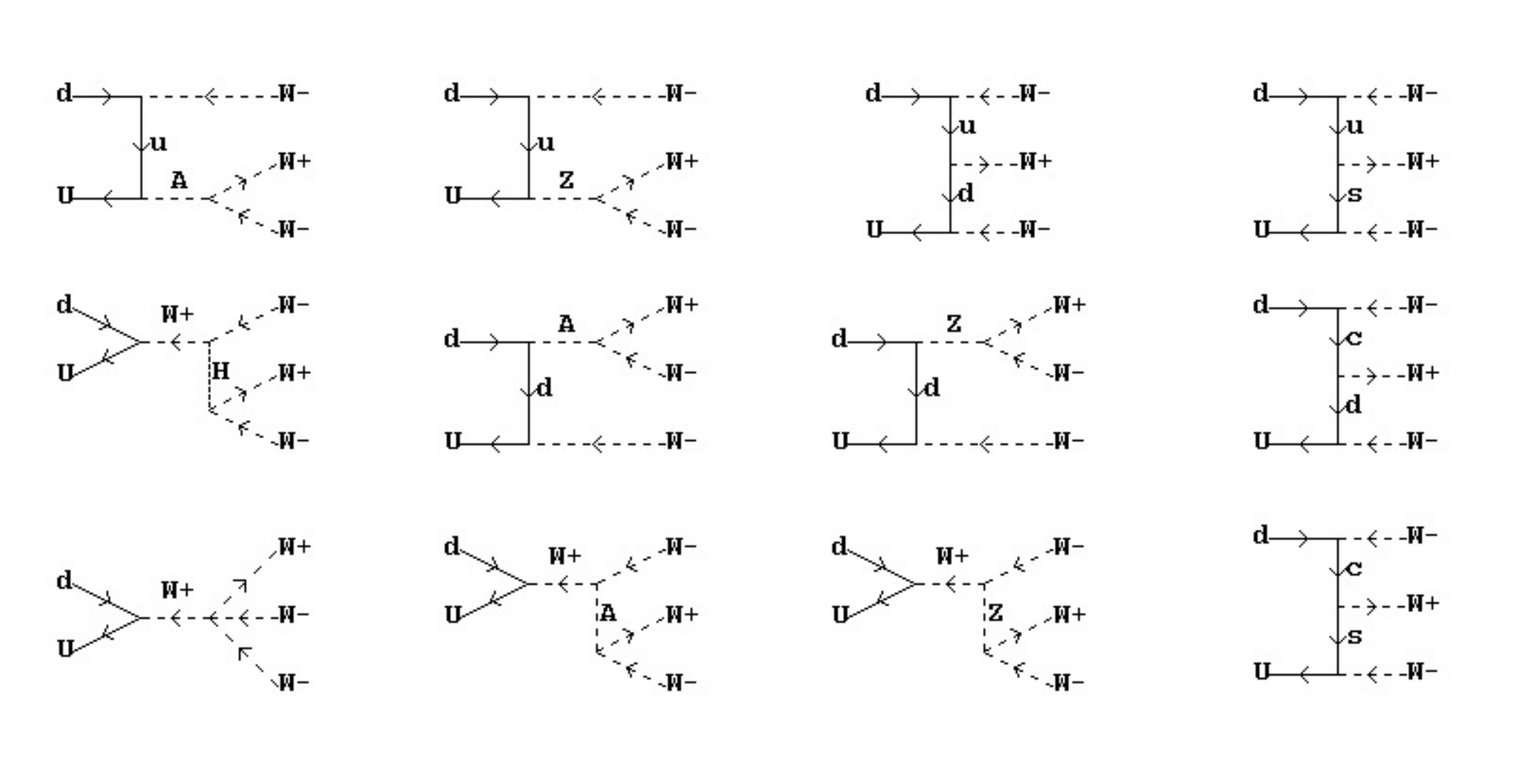}
\caption{Diagrams for subprocess $U+d \to W^+W^-W^-$.}
\label{fig:W--}
\end{figure}

 Leading order diagrams for processes~(\ref{W+},\ref{W-}) for the main channels with initial states $u D$ and $d U$ are shown at Figures \ref{fig:W++},\ref{fig:W--}. Analogous sets of diagrams are also taken into account for initial states
 $u S,\,c D,\, c S,\, s U, d C, s C$. Calculations are performed in the framework of CompHEP package for evaluation of Feynmann diagrams, integration over multi-particle phase space and event generation~\cite{CompHEP}.
\bigskip

Results of calculations with the use of CompHEP package of diagrams for  processes~(\ref{W+}, \ref{W-}) are shown in Table 1, where we estimate also event numbers of processes~(\ref{W+dec}, \ref{W-dec})  for two values of the integral luminosity.
The first one is already achieved $L\,=\,36.4\,fb^{-1}$ and the second one is $L\,=\,130\,fb^{-1}$,
which presumably is expected for the overall Run II resulting value. We present results in dependence on the module of parameter $\lambda$. Calculations show that the results in the Table as well as in the
subsequent ones in fact do not depend on the sign of $\lambda$.
\bigskip
\begin{center}
Table 1.\\
\end{center}
LO cross sections of three-boson production $W^+ W^- W^\pm$ and event numbers with signature (\ref{W+dec}, \ref{W-dec}) for two values of the integral luminosity  for  $\sqrt{s}=13\,TeV$  at the
LHC for six values of $|\lambda|$.
\begin{center}
\begin{tabular}[l]{|||c|c|c|c|c|c|c|c|||} \hline
$|\lambda|$ &0.011 & 0.0085 & 0.006 & 0.0045 & 0.003 & 0.002 & 0.000\\
\hline
 $\sigma\, fb $& 1419.8 & 884.3 & 501.9 & 320.9 & 212.9 & 158.9 & 116.3 \\
\hline
$N_{36.4}$  & $327\pm18$ & $203\pm14$ & $115\pm11$ &$74\pm 9$ & $49\pm7$ & $37\pm6$ & $27\pm5$\\
 \hline
$\Delta N_{36.4}$  & $300\pm18$ & $177\pm14$ & $88\pm11$ &$47\pm9$ & $22\pm7$ & $10\pm6$ & 0 \\
 \hline
 $N_{130}$  &$1166\pm34$ &$726\pm27$ &$412\pm20$ & $263\pm16$ & $175\pm13$ & $130\pm11$ & $95 \pm 10$\\
\hline
 $\Delta N_{130}$  &$1071\pm34$ &$631\pm27$ &$317\pm20$ & $168\pm16$ & $80\pm13$ & $35\pm11$ &  0\\
 \hline \hline
 \end{tabular}
\end{center}

We could compare our estimates with recent work by ATLAS~\cite{ATLAS19}, where the first data on
triple $W$ production at $\sqrt{s} = 13\,TeV$ are presented. For the process under the discussion
(\ref{W+},\ref{W-}) the following result was obtained
\begin{equation}
\sigma_{WWW} = 0.68^{+0.23}_{-0.21}\,pb.\label{ATLAS19}
\end{equation}
Comparing this result with calculations of cross-sections presented at Table 1, we estimate the possible value of $|\lambda|$, which corresponds to result~(\ref{ATLAS19})
\begin{equation}
|\lambda|\,=\,0.0073^{+0.0026}_{-0.0030}\,.\label{res1}
\end{equation}
The result is safely inside limitations~(\ref{lambda17}, \ref{lambda19}).

Let us consider also another process of the three-boson production
 \begin{equation}
p+p \to W^+\,W^-\,Z\,+\,no\,\,jet\,; \label{WWZ}
\end{equation}
We have also to take into account characteristic leptonic decays of the bosons, which lead to the following
combinations under the condition, that an invariant mass of the two last oppositely charged leptons
corresponds to the $ Z $ boson:
\begin{eqnarray}
& &W^+\,W^-\,Z\,\to \mu^+ \,e^-\,  e^+\, e^- + neutrinos;\nonumber \\
& &W^+\,W^-\,Z\,\to \mu^- \,e^+\,e^+\, e^- + neutrinos;\label{WZ-dec} \\
& &W^+\,W^-\,Z\,\to \mu^+ \,e^-  \mu^+\, \mu^- + neutrinos;\nonumber \\
& &W^+\,W^-\,Z\,\to \mu^- \,e^+\,\mu^+\, \mu^- + neutrinos.\nonumber
\end{eqnarray}
Calculations lead to results shown in Table 2.
\newpage
\begin{center}
Table 2.\\
\end{center}
Cross sections of process $W^+ W^- Z$ in the leading order and event numbers with
signature~(\ref{WZ-dec})  for
two values $L$ at LHC,  $\sqrt{s}=13\,TeV$ at the LHC for six values of $|\lambda|$.
\begin{center}
\begin{tabular}[l]{|||c|c|c|c|c|c|c|c|||
} \hline
$|\lambda|$ &0.011 & 0.0085 & 0.006 & 0.0045 & 0.003 & 0.002 & 0.000\\
\hline
 $\sigma\, fb $ & 794.0 &  509.0 & 298.3 & 206.6 & 140.4 & 111.4 & 87.9 \\
\hline
$ N_{36.4}$  & $ 45\pm7 $ & $ 29\pm5 $ & $17\pm4$ & $ 12\pm 3 $ & $ 8\pm3 $ & $ 6\pm3$ & $ 5\pm2$\\
 \hline
$ \Delta N_{36.4} $  & $ 40\pm7 $ &  $24\pm5 $ & $ 12\pm4 $ & $ 7\pm3 $ & $ 3\pm3 $ & $ 1\pm3 $ & 0 \\
 \hline
 $ N_{130} $  &$ 162\pm13$ &$ 104\pm10 $ &$ 61\pm8 $ & $ 42\pm6 $ & $ 29\pm5 $ & $ 23\pm5 $ & $ 18 \pm 4 $\\
\hline
 $ \Delta N_{130} $  &$ 144\pm13 $ &$ 86\pm10 $ &$ 43\pm8 $ & $ 24\pm6 $ & $ 11\pm5 $ & $ 5\pm5 $ &  0\\
\hline \hline
 \end{tabular}
\end{center}

From result~\cite{ATLAS19} for reaction $ p+p\to W^+ W^- Z$
\begin{equation}
\sigma_{WWZ} = 0.49^{+0.20}_{-0.18}\,pb;\label{ATLAS19Z}
\end{equation}
we have with calculations, presented in Table 2
\begin{equation}
|\lambda|\,=\,0.0083^{+0.0028}_{-0.0026}\,.\label{res2}
\end{equation}
Let us quote also the estimate for possible value of parameter $|\lambda|$, which was obtained in  work~\cite{AZPL17} from data for process
\begin{equation}
 p + p \to \bar t t H +X\,;\label{ttH}
\end{equation}
\begin{equation}
 |\lambda|\,=\,0.0057^{+0.0039}_{-0.0028}.\label{ttHl}
\end{equation}
All these estimates do not contradict restrictions~(\ref{lambda17}, \ref{lambda19}, \ref{lambdaWZ}).
Results~(\ref{res1}, \ref{res2}, \ref{ttHl}) are mutually consistent, and each one do not contradict to the zero value for $\lambda$.
 In our approach effects in these three processes have the same origin and are connected with
anomalous interaction~(\ref{FFF}) with effective coupling constant (\ref{Glam}), which is defined by parameter~$\lambda$. In the framework of this hypothesis we may average out three results~(\ref{res1}, \ref{res2}, \ref{ttHl}), and come to preliminary rough estimate
\begin{equation}
|\lambda|\,=\,0.0071\,{\pm 0.0026}.\label{lambdal}
\end{equation}
This estimate is also consistent with the zero effect. However with increasing integral luminosity we may hope to achieve decisive results on a possibility of $\lambda$ being of the order of magnitude of the mean value of estimate~(\ref{lambdal}).

It would be instructive to compare results being shown in Table 1 with those, which could be obtained in a course of studying of the same process at $\sqrt{s} = 8\,TeV$. In doing this we would present estimates for integral luminosity $L =19\,fb^{-1}$, which was achieved
in experiment~\cite{CMS17} leading to limitations~(\ref{lambda17}). Results are shown in Table~2.
By comparing Table 1 and Table 2 we see, that the effect in reaction under the discussion for
conditions, which correspond to $\sqrt{s}=8\,TeV$ with the actual luminosity, is practically absent, while for $\sqrt{s}=13\,TeV$ one may hope for essential improving of data~(\ref{lambda17},~\ref{lambda19}) even for integral luminosity $L\,=\,36\,fb^{-1}$.
The integral luminosity $L\,=\,130\,fb^{-1}$, which presumably might be achieved in RUN II, provides a  quite effective tool for dealing with effects of small $|\lambda|$ even
as small as $|\lambda| \simeq 0.002$.
\newpage
\begin{center}
Table 3.\\
\end{center}
LO cross sections of three-boson production $W^+ W^- W^\pm$ for  $\sqrt{s}=8\,TeV$  at the
LHC for six values of $|\lambda|$ and predicted number of events for $L = 19\,fb^{-1}$.
\begin{center}
\begin{tabular}[l]{|||c|c|c|c|c|c|c|c|||
} \hline
$|\lambda|$ &0.011 & 0.0085 & 0.006 & 0.0045 & 0.003 & 0.002 & 0.000\\
\hline
 $\sigma\, fb $& 124.7 & 100.2 & 75.5 & 66.5 & 61.7 & 56.7 & 49.4 \\
\hline
$N_{19}$  & $15\pm4$ & $12\pm3$ & $9\pm3$ &$8\pm 3$ & $7\pm3$ & $7\pm3$ & $6\pm2.5$\\
 \hline
$\Delta N_{19}$  & $9\pm4$ & $6\pm3$ & $3\pm3$ &$2\pm3$ & $1\pm3$ & $1\pm3$ & 0 \\
 \hline \hline
 \end{tabular}
\end{center}

We would also compare our results with possibilities provided by another process
\begin{equation}
p+p \to W^\pm\,Z\,\gamma\,+\,no\,\,jet\,; \label{WZgam}
\end{equation}
which also could be used to define parameter $\lambda$ more exactly. Again with application of the CompHEP~\cite{CompHEP}, we calculate corresponding cross sections of the reaction~(\ref{WZgam}) in dependence on a minimal $\gamma$ transverse momentum for $|\lambda|$ in the same region. In estimating
of event number we take into account the following leptonic signatures provided an invariant mass of the two last oppositely charged leptons
corresponds to the $ Z $ boson:
\begin{eqnarray}
& &W^\pm\,Z\,\gamma\to \mu^\pm\,e^+\, e^-\gamma + neutrino;\nonumber \\
& &W^\pm\,Z\,\gamma\to e^\pm \,\mu^+\,\mu^-\,\gamma + neutrino;\label{WZG-dec}\\
& &W^\pm\,Z\,\gamma\to \mu^\pm\,\mu^+\, \mu^-\,\gamma + neutrino;\nonumber \\
& &W^\pm\,Z\,\gamma\to e^\pm \,e^+\,e^-\,\gamma + neutrino.\nonumber
\end{eqnarray}
\begin{center}
Table 4.\\
\end{center}
Leading order cross-section of  $W^\pm Z \gamma$ production for $\sqrt{s}=13\,TeV$ at the LHC with $p_T^\gamma> 50\,GeV$ and estimates for event numbers for two values of an integral luminosity in dependence of value
 $|\lambda|$.
\begin{center}
\begin{tabular}[l]{|||c|c|c|c|c|c|c|c|||
} \hline
$|\lambda|$ &0.011 & 0.0085 & 0.006 & 0.0045 & 0.003 & 0.002 & 0.000\\
\hline
 $\sigma\, fb $ &  201.9 & 130.9 & 76.8 &  53.4 & 36.8 & 29.4 & 23.3 \\
\hline
$ N_{36.4}$  & $ 125\pm11 $ & $ 81\pm9 $ & $ 48\pm 7$ & $ 33\pm 6 $ & $ 23\pm5 $ & $ 18\pm4 $ & $ 14\pm4$\\
 \hline
$ \Delta N_{36.4}$  & $ 111\pm11 $ & $ 67\pm9 $ & $ 34\pm 7 $ &$ 19\pm6 $ & $ 9\pm5 $ & $ 4\pm4 $ & 0 \\
 \hline
 $ N_{130}$  &$ 448\pm21 $ &$ 290\pm17 $ &$ 170\pm13 $ & $ 118\pm11 $ & $ 82\pm9 $ & $ 65\pm8 $ & $ 52 \pm 7 $\\
\hline
 $ \Delta N_{130}$  &$ 396\pm21 $ &$ 238\pm17 $ &$ 118\pm13 $ & $ 66\pm11 $ & $ 30\pm9 $ & $ 13\pm8 $ &  0\\
\hline \hline
\end{tabular}
\end{center}

\begin{center}
Table 5.\\
\end{center}
Leading order cross-section of  $ W^\pm Z \gamma $ production for $ \sqrt{s}=13\,TeV $ at the LHC  with $ p_T^\gamma> 100\,GeV $ and estimates for event numbers for two values of an integral luminosity in dependence of value
 $ |\lambda| $.
\begin{center}
\small{
\begin{tabular}[l]{|||c|c|c|c|c|c|c|c|||
} \hline
$ |\lambda|$ &0.011 & 0.0085 & 0.006 & 0.0045 & 0.003 & 0.002 & 0.000\\
\hline
 $\sigma\, fb $ & 184.1 & 114.7 & 60.6 & 38.0 & 21.3 & 14.1 & 8.3 \\
\hline
$ N_{36.4}$  & $ 114\pm11 $ & $ 71\pm9 $ & $ 38\pm6 $ & $ 24\pm 5 $ & $ 13\pm4 $ & $ 9\pm3$ & $ 5\pm2 $\\
 \hline
$ \Delta N_{36.4}$  & $ 109\pm11 $ & $ 66\pm9 $ & $ 33\pm6 $ &$ 19\pm5 $ & $ 8\pm4 $ & $ 4\pm3 $ & 0 \\
 \hline
 $ N_{130}$  &$ 408\pm20 $ &$ 254\pm16 $ &$ 134\pm12 $ & $ 84\pm9 $ & $ 47\pm7 $ & $ 31\pm6 $ & $ 18 \pm 4 $\\
\hline
 $ \Delta N_{130} $  &$ 390\pm20 $ &$ 236\pm16 $ &$ 116\pm12 $ & $ 66\pm9 $ & $ 29\pm7 $ & $ 13 \pm6 $ &   0\\

 \hline \hline
 \end{tabular}}
\end{center}
With account of leptonic branching ratios of $W,\,Z$ we see from Table 4 and Table 5, that while
processes~(\ref{W+}, \ref{W-}) with decays (\ref{W+dec}, \ref{W-dec}) are much more promising for
looking for small $\lambda$ effects than process (\ref{WZgam}) with decays (\ref{WZG-dec}), the last
one (\ref{WZgam}) nevertheless could give additional information on the problem of non-perturbative
contributions, especially with cut-off $p_T^\gamma = 100\,GeV$.

\section{Conclusion}
\label{sec:conclusion}
The study of processes, which was discussed above would provide essential improvement of restrictions for value of $\lambda$, in particular, give decisive  answer for a possibility (\ref{lambdal})
\begin{equation}
|\lambda| \simeq 0.007.\label{007}
\end{equation}
Especially we would draw attention to processes
\begin{eqnarray}
& &p\,p\,\to W^+ W^+ W^-;\nonumber\\
& &p\,p\,\to W^- W^- W^+;\label{reaction}
\end{eqnarray}
and look for the low background experimental signature
\begin{eqnarray}
& &p\,p\,\to \mu^+ \mu^+ e^- + invisible;\nonumber\\
& &p\,p\,\to \mu^- \mu^- e^+ + invisible;\label{signature}\\
& &p\,p\,\to e^+ e^+ \mu^- + invisible;\nonumber\\
& &p\,p\,\to e^- e^- \mu^+ + invisible.\nonumber
\end{eqnarray}
in which according to Table 1 the effect essentially exceeds the SM
predictions.

We would draw attention to an importance of searches for non-perturbative effects in
the electro-weak interaction. Just anomalous interaction of $W$-bosons (\ref{FFF}) in case of confirmation of its existence would open a new region of investigations in the field of the non-perturbative electro-weak physics.

\end{document}